# Orbiton-Phonon coupling in Ir$^{5+}$(5$d^4$) double perovskite Ba$_2$YIrO$_6$

Birender Singh[1], G. A. Cansever[2], T. Dey[2], A. Maljuk[2], S. Wurmehl[2,3], B. Büchner[2,3] and Pradeep Kumar[1*]

[1]School of Basic Sciences, Indian Institute of Technology Mandi, Mandi-175005, India

[2]Leibniz-Institute for Solid State and Materials Research, (IFW)-Dresden, D-01171 Dresden, Germany

[3]Institute of Solid State Physics, TU Dresden, 01069 Dresden, Germany

## ABSTRACT:

Ba$_2$YIrO$_6$, a Mott insulator, with four valence electrons in Ir$^{5+}$ $d$-shell (5$d^4$) is supposed to be non-magnetic, with J$_{eff}$ = 0, within the atomic physics picture. However, recent suggestions of non-zero magnetism have raised some fundamental questions about its origin. Focussing on the phonon dynamics, probed via Raman scattering, as a function of temperature and different incident photon energies, as an external perturbation. Our studies reveal strong renormalization of the phonon self-energy parameters and integrated intensity for first-order modes, especially redshift of the few first-order modes with decreasing temperature and anomalous softening of modes associated with IrO$_6$ octahedra, as well as high energy Raman bands attributed to the strong anharmonic phonons and coupling with orbital excitations. The distinct renormalization of second-order Raman bands with respect to their first-order counterpart suggest that higher energy Raman bands have significant contribution from orbital excitations. Our observation indicates that strong anharmonic phonons coupled with electronic/orbital degrees of freedom provides a knob for tuning the conventional electronic levels for 5$d$-orbitals, and this may give rise to non-zero magnetism as postulated in recent theoretical calculations with rich magnetic phases.

[*] Corresponding author: email id: pkumar@iitmandi.ac.in

# 1. INTRODUCTION

Recently, the interest in transition metal oxides with $4d/5d$ valence electrons has grown manifold. As one moves from $3d$ to $5d$ systems, Coulomb interaction (U) is expected to be weaker because of the larger radial extent of $5d$ orbitals. On the other hand spin-orbit coupling (SOC) follows the opposite behaviour. In systems like $5d$ iridium based oxides a peculiar situation arises where SOC competes with U and this competition along with interaction with crystal field splitting ($\Delta_{CF}$), intersite hopping ($t_{ij}$), Hund's coupling ($J_H$) and superexchange gives rise to rich ground states and quasi-particle excitations [1-6] and the most interesting effect of the fierce competition between these different interactions is reflected via the observations of $J_{eff} = 1/2$ Mott state [7-8].

The focus in the literature so far has been concentrated on $Ir^{4+}$ ($5d^5$) systems, where the physics of these systems is mainly believed to be driven by $J_{eff} = 1/2$ state [9]. However, $Ir^{5+}$ ($5d^4$) systems have not been explored much, where the strong SOC is expected to gives rise to non-magnetic singlet ground state with $J_{eff} = 0$. Infact, such a picture of effective zero total angular momentum has been invoked to explain the non-magnetic nature of $Ir^{5+}$ systems such as $NaIrO_3$ [10]. Interestingly, the recent suggestion of non-zero magnetism in $Ir^{5+}$ double perovskites $A_2YIrO_6$ (A = Ba, Sr) [11-14] have raised some fundamental questions. In particular, whether $J_{eff} = 0$ picture needs to be revisited. Is it affected by the presence of some non-cubic crystal field ($\delta_{CF}$) or because of extended $5d$ states, as these states are expected to generate a strong coupling between magnetic, electronic as well as phononic degrees of freedom (DOF), $J_{eff}$ picture breaks down.

In $Ba_2YIrO_6$, the $Ir^{5+}$ is octahedrally coordinated with oxygen ions. It results into splitting of $d$-orbitals into doubly and triply degenerate $e_g$ ($d_{z^2}, d_{x^2-y^2}$) and $t_{2g}$ ($d_{xy}, d_{xz}, d_{yz}$) levels, respectively [15], which leads to high spin (S =1) state with all four electrons in $t_{2g}$ orbitals (shown schematically in Fig. 1 (ii)). Degeneracy of $e_g / t_{2g}$ levels may be further lifted if symmetry of $Ir^{5+}$ ions ($O_h$) is lowered. Non-cubic crystal field ($\delta_{CF}$), which may arise if apical oxygen atoms are pushed toward Ir-ion, interactions with nearest octahedra or strong anharmonic phonons associated with the vibrations of $IrO_6$ octahedra, may further lift the degeneracy of $t_{2g}$ orbitals then $d_{xy}$ orbital is stabilized with lower energy as compared to $d_{xz}$, $d_{yz}$ with symmetry reduced to $D_{4h}$. Infact, apical oxygen atoms in case of $Ba_2YIrO_6$ are reported to be closer to the Ir ion than basal oxygens [16], and therefore, it may provide a fertile ground for the lifting of $t_{2g}$ degeneracy as suggested schematically in figure 1 (iii). But the strong SOC may result into different configuration of $t_{2g}$ levels (see figure 1 (iv) ). We note that estimated strength of $\delta_{CF}$, in other $5d$-Ir based system without considering the phononic DOF, is ~ 150-600 meV and SOC is ~ 300-500 meV [17], therefore one need to consider the effect of non-cubic crystalline fields in iridium based systems especially that of generated by phonons, quanta of lattice vibrations. The probability of this configuration (as shown in Fig. 1 (iii)) is non-zero and will depend on the strength of non-cubic crystal field.

So far, the focus in the literature has been limited to exploring the intricate coupling between SOC, U, $J_H$, tunnelling parameter and superexchange [1-6]. The role of phononic DOF has been completely unexplored. Therefore, it becomes inevitable to probe the role of phonons and their potential coupling with electronic/orbital DOF. In this paper, we have undertaken such a study and have explored the role of phonons in $Ba_2YIrO_6$ using Raman scattering. As Raman is a very powerful technique to probe quasi-particle excitations and their coupling, any potential coupling of phonons with other DOF will be reflected via renormalized phonon self-energy parameters, which may be observed in Raman scattering.

Here, we report the inelastic light scattering studies on $Ba_2YIrO_6$, in a wide temperature (80-400 K) and spectral range of 50-1700 $cm^{-1}$. The strong renormalization of the modes associated with $IrO_6$ octahedra (i.e. S10-S15) and their second-order counterpart evince the non-zero contribution of the non-cubic crystal field ($\delta_{CF}$), which may potentially explain the experimental observation of non-zero magnetism in this 'expected' non-magnetic system. The anomalous temperature and incident photon energy dependence of high energy Raman modes (i.e. S16-S19), in the spectral range of 1100-1500 $cm^{-1}$, have their origin in orbital excitations. Our observations, discussed in details in sections below, clearly suggest an intricate role of phononic DOF, which may play an important role in explaining the rich magnetic phases. Our results cast a crucial light on the role of phononic DOF in these $5d^4$ double perovskite systems and suggest that phonons should be treated at par with the electronic DOF to understand the underlying exotic properties of these systems.

## 2. METHODS

### 2.1. Experimental Details

Unpolarised micro-Raman measurements were performed on single crystals of $Ba_2YIrO_6$, with typical size of 0.5*0.5*0.5 $mm^3$, in backscattering geometry using laser wavelength of 532 nm and Raman spectrometer (Labram HR-Evolution) coupled with a Peltier cooled CCD. Laser power at the sample was kept very low (~ 1 mW) to avoid any heating of the sample. Temperature variation was done from 80 K to 400 K, with a temperature accuracy of 1K using continuous flow liquid nitrogen cryostat (Linkam Scientific).

### 2.2. Computational Details

First principles calculations were performed utilizing the ultrasoft pseudopotential plane-wave [18] approach within the framework of the density functional theory (DFT) as implemented in QUANTUM ESPRESSO [19]. The dynamical matrix and phonons at gamma point (q = 0,0,0) were calculated using Perdew-Burke-Ernzerhof generalized gradient approximation [20] as exchange-correlation functional along with linear response approach within density functional perturbation theory [21]. The plane wave cutoff energy of 30 Ry and the charge density cutoff of 350 Ry was used. The numerical integration of Brillouin zone (BZ) were performed with 4 x 4 x 4 Monkhorst-Pack [22] k-point mesh. The plane wave cutoff and k-points used in calculations were set after performing their respective convergence test.

## 3. Results and Discussions

### 3.1. Raman Scattering from Phonons

$Ba_2YIrO_6$ crystallizes in cubic (space group $Fm\overline{3}m$) as well as monoclinic (space group $P2_1/n$, with $\beta = 90.039°$) double perovskite type structure [11-12,16,23]. Numbers of first-order Raman active modes in cubic and monoclinic structure with irreducible representations are $\Gamma_R = A_{1g} + E_g + 2T_{2g}$ and $\Gamma_R = 12A_g + 12B_g$, respectively [24]. Figure 2 shows the Raman spectra at 80 K, revealing 19 modes labeled as S1 to S19 in the spectral range of 60-1700 cm$^{-1}$. Spectra are fitted with a sum of Lorentzian functions to extract the peak frequency, linewidth and area under the curve; the individual modes are shown by thin lines, and the resultant fit is shown by a thick line. To identify the phonon modes observed in Raman spectra, we calculated phonon frequencies at gamma point ($q$ = 0,0,0) for both experimental as well as well as theoretical lattice parameters in monoclinic symmetry. The lattice constants used are: experimental, $a$ = 5.9028 Å, $b$ = 5.9029 Å, $c$ = 8.3500 Å; and theoretical, $a$ = 5.8164 Å, $b$ = 5.8167 Å, $c$ = 8.2271 Å. Our first-principle theoretical estimates of optimized lattice constants are ~ 1.5 % smaller than experimental lattice constants. Calculated phonon frequencies for theoretical lattice parameter are in good agreement with the experimental observed values at 80K (see Table-I). We made a comparison between calculated phonon frequency for the experimental lattice parameters and the observed frequency at 80 K as:

$$\left. |\bar{\omega}_r| \right|_{\%} = \frac{100}{N} \sum_i \left| \frac{\omega_i^{cal} - \omega_i^{exp}}{\omega_i^{exp}} \right|,$$ Where $i = 1, 2 - - - -, N$ is the number of Raman active modes (15 here). $\left. |\bar{\omega}_r| \right|_{\%}$ is the average of the absolute relative differences in percentage. The estimated relative average in the mode frequencies is found to be ~ 3.5 % only and is within the typical error expected in the DFT based calculations. Based on our first-principle DFT based calculation, we have assigned modes below ~ 800 cm$^{-1}$, i.e. modes S1-S15, as first-order phonon modes and high frequency modes, i.e. S16-S19, have been assigned as second order modes coupled with orbital excitations.

## 3.2. Temperature dependence of the first-order phonons

Figure 3 shows the temperature dependence of the self-energy parameters (i.e. mode frequencies and full width at half maxima (FWHM)) of the prominent first-order phonon modes S2, S4, S8 and S9 in the spectral range of 50-500 cm$^{-1}$. The following observations can be made : (i) most striking and interesting is the redshift, i.e. mode frequency ($\omega$) decreases with decreasing temperature, of the modes S2 and S8. However, their linewidths (i.e. FWHM) shows normal temperature dependence i.e. FWHM decreases with decreasing temperature. Frequency of mode S2 shows a small drop around 250 K, which may have its origin in subtle local distortion. (ii) Mode frequencies and FWHM of mode S4, S9 and weak modes S1, S3 (not shown here) shows normal temperature dependence, i.e. mode frequency increases and FWHM decreases with decreasing temperature as expected because of anharmonic effects. (iii) FWHM of the modes S2, S4, and S8 decreases by ~ 100-200 % at 80 K with respect to its value at 400 K, and other modes are much sharper at low temperature, suggesting the absence of disorder in the system with respect to the mixing of Y and Ir.

Figure 4 shows the temperature dependence of the prominent first-order modes in the spectral range of 500-900 cm$^{-1}$. The following observations can be made: (i) Temperature dependence of all the modes shows normal behaviour i.e. peak frequencies ($\omega$) and FWHM increase and decreases, respectively, with decreasing temperature. Interestingly, mode S10-S12 and S15 show a large softening ~ 1.4 - 2 % in the temperature range of 80-400 K. (ii) With decreasing temperature FWHM decreases by ~ 100-400 % suggesting absence of disorder. (iii) Significant transfer of spectral weight (see inset (a) of Fig. 2) from mode S10 to S11 as temperature is lowered hinting towards the reorientations of the molecular units associated with these modes (i.e. IrO$_6$ octahedra). As Raman tensor is very sensitive to the molecular orientation in space, any change in the molecular orientation will be reflected in the corresponding change in the intensity of the Raman bands associated with it. Similar transfer of spectral weight has been observed in H-bonded systems associated with the molecular reorientations [25-26].

We will now discuss the temperature dependence of these modes. A solid may be approximated by assuming that atoms are connected via springs, with spring constant $k$ and within the harmonic approximation mode frequency ($\omega_o$) is proportional to $\sqrt{k}$. As the temperature is lowered, bond length shrink or stiffness of the spring increases, therefore one may expect that mode frequency ($\omega$) should increase with decreasing temperature. At the same time with decreasing temperature, anharmonic interaction (i.e. $U_{anhar.}(r) = gr^3 + --- \equiv \beta[a^+a^+a + a^+aa] + ---$; where $a^+$, $a$ are phonon creation and annihilation operator, respectively, and cubic term will contribute only in second order perturbation theory) decreases and as a result phonon lifetime ($\alpha \frac{1}{FWHM}$) increases owing to reduced interactions with other quasi-particles. In order to understand the effect of anharmonicity, we have fitted the self-energy parameters (i.e. $\omega$ and FWHM ($\Gamma$)) based on cubic anharmonic model where an optical phonon decay into two-phonons [27] of equal energy giving the temperature dependence of $\omega(T)$ and $\Gamma(T)$ as: $\omega(T) = \omega_0 + A\{1 + [2/(e^x - 1)]\}$ and $\Gamma(T) = \Gamma_0 + C\{1 + [2/(e^x - 1)]\}$, where $x = \frac{\hbar\omega_0}{2k_BT}$ and A, and C are self-energy constant. $\omega_0$ and $\Gamma_0$ are the mode frequency and linewidth at zero Kelvin. All the modes can be fitted reasonably well using the above model (see solid lines in figure 3 and 4). List of constants are given in Table-I. We note that constant A associated with modes S10-S15 is very high reflecting strong anharmonic nature of these modes.

Most interesting observation is the redshift of modes S2, S6-S8, (S6 and S7 are not shown) which is opposite to the conventional behaviour of phonons as described above. This may be understood by invoking two possible mechanisms: in first picture coupling of phonons with magnetic order (spin-phonon coupling) may soften the phonon with decreasing temperature, and second is that the strong phonon-phonon interaction may also soften phonon modes. The change in phonon frequency due to spin-phonon coupling ($\Delta\omega_{Sp-Ph}$) may arise due to modulation of the exchange integral by the phonon amplitude and/or by involving change in the Fermi surface by spin waves provided phonon couple to that part of the Fermi surface. The contribution from spin-phonon coupling may be given as [28-29] $\Delta\omega_{Sp-Ph} = \lambda < \vec{S}_i \cdot \vec{S}_j >$, where $< \vec{S}_i \cdot \vec{S}_j >$ is the spin correlation function and $\lambda$ is the spin-phonon coupling constant which is phonon dependent and which can be positive as well as negative. But spin-phonon coupling is expected to be active only at temperature where the magnetic correlation exists. However, Ba$_2$YIrO$_6$ is found to be in paramagnetic state down to 2 K [11], but a signature of long range magnetic ordering has been reported at ~ 1.5K [12] therefore contribution of spin DOF to the redshift of these phonons is expected to be marginal. It will be interesting to probe the role of spin DOF on phonon dynamics using theoretical calculations as well as first-principle calculations via mapping phonon density of states in different possible magnetic ground states.

The red shift of these modes may also originate from strong phonon-phonon anharmonic interactions. In general, self-energy constant (A) is negative i.e. mode frequency decreases with increasing temperature, which is referred in literature as normal behaviour. However, it may be positive, in which case mode frequency may increase with increasing temperature, called anomalous behaviour. The self-energy of a phonon has two parts, real ($\Sigma_r$) and imaginary ($\Sigma_i$). Imaginary part is related to the phonon lifetime i.e. $\Delta\Gamma(T)$ $\alpha$ $\Sigma_i = |V(\omega)|^2 \rho_2(\omega)$, where V is the anharmonic coupling constant and $\rho_2(\omega)$ is two-phonon density of states (integrated over all decay channel with constraints $\omega = \omega_1 + \omega_2$ and $\vec{q} = 0 = \vec{q}_1 + \vec{q}_2$), and the real part is associated with phonon frequency as $\Delta\omega(T)$ $\alpha$ $\Sigma_r = \frac{2|V|^2}{\pi} \int_0^\infty \frac{\rho_2(\omega')}{\omega^2 - \omega'^2} d\omega'$, assuming that V is independent of frequency. Now, sign of this integral will determine the sign of self-energy constant (A), and generally it is negative because terms with $\omega' > \omega$ dominate. However, if reverse happens i.e. where peak in two-phonon density of states is lower than $\omega$ then integral may be positive and it may be reflected in the red shift of phonon frequency. Therefore, sign of self-energy constant can be positive or negative depending on whether the two-phonon density of states has a maximum at a frequency lower or higher than the phonon frequency [30]. We note that the self-energy constant (A) associated with these modes is very high similar to the modes S10-S15 in addition to their red-shifting with lowering temperature reflecting strong anharmonic nature of these modes.

### 3.3. Orbiton-phonon coupling and second order phonons

Looking at the frequency range (1100-1500 cm$^{-1}$) (see Fig. 2) of these high energy Raman bands (S16-S19), one would expect them to be second-order phonons as their frequency range falls within the double of the first-order phonon frequency. However, upon careful detailed analysis of their temperature and incident photon energy ($\omega_L$) behaviour, they reveal non-zero contribution of other quasi-particles excitations, namely orbitons (or orbital excitations), which are well studied in case of 3$d$ oxides [31-36]. We know that 3$d$ systems have very active orbital DOF in addition to spin degrees of freedom owing to very weak SOC and Jahn-Teller mechanism generate strong coupling between phonons and orbital degrees of freedom [37]. On the other hand in case of 5$d$ systems, strong SOC is expected to quench orbital DOF and as a result Jahn-Teller mechanism is suppressed along with coupling of phonons with orbital DOF. However, counter to this conventional belief it has been argued, in recent theoretical as well as experimental reports [38-41], that even in case of Ir (5$d$) based systems Jahn-Teller mechanism does indeed effect $t_{2g}$ manifold despite strong SOC, attributed to the fact that phonon driven Jahn-Teller mechanism only couples to the orbital DOF. It has been also suggested that at moderate intersite hopping by electrons, SOC is not strong enough to quench orbital dynamics completely and this results into emergence of non-zero moments attributed to the entanglement between spin and orbital DOF mediated by lattice. Also, in a recent study on similar systems via RIXS measurements [41], SOC (350-400 meV) is estimated to be of comparable strength as J$_H$ (~ 250 meV). This convey that SOC is not the only driving parameters, as oppose to the conventional belief, to control the physics of these systems and orbital degrees of freedom should be treated on equal footing.

Therefore, probing the role of orbital degrees of freedom becomes very important to understand the underlying microscopic phenomena in these systems. As the assignment of second-order Raman bands specific to a particular first-order phonon is not straight forward, since second-order Raman scattering involves the phonons over the entire Brillouin zone with major contribution from region of higher density of states, generally all possible contributions results in broad second-order modes in comparison to first-order modes. Also, the peak frequency of second-order bands are not necessarily double of those of the first-order phonons at the gamma point. As a starting point, we have assigned the second-order modes S16-S19 as overtone of first-order modes S10-S13, respectively. We note that, these second-order modes may also be assigned as possible combinations of different first-order modes, e.g. S19 may be assigned as combination of S15, S12; and S14, S12.

Figure 5 (b, c, e and f) shows the temperature evolution of the peak frequency and FWHM of modes S16-S19. The following observations can be made: (i) peak frequency and FWHM of mode S16, S17 shows normal behaviour. (ii) Interestingly, the peak frequency of mode S18 and S19 shows anomalous behaviour, and (iii) linewidth of mode S19 shows anomalous broadening with decreasing temperature. The anomalous behaviour of these modes, large red-shift of mode S18 by ~ 40 $cm^{-1}$ and linewidth increase by ~ 50% of S19, clearly suggests that these modes are strongly coupled with other excitations. Top panel (i.e. Fig. 5(a) and 5(d)) shows the intensity ratio of second-order Raman bands w.r.t. their first-order counterparts. As the second-order modes become weak at high temperature, so as a qualitative characterization of their intensity ratio, we have plotted the intensity ratio of Raman band with centring around 1150 $cm^{-1}$ (S16 and S17) w.r.t. their corresponding first-order modes (S10 and S11) i.e. $I_{16+17}/I_{10+11}$ (see Fig. 5a). Similarly, we have plotted the intensity ratio (i.e. $I_{18+19}/I_{12+13}$) for band with centring around 1450 $cm^{-1}$ (i.e. S18 and S19), which increases with decreasing temperature (see Fig. 5d), and this increase in intensity may be understood invoking resonant Raman scattering. Within the resonance picture, the increase in intensity with decreasing temperature arises from an increase in the lifetime of the resonant excitation at low temperature [42] and exactly at the resonance, the Raman intensity ratio of second to first-order phonons is proportional to the square of the lifetime of the excited state. At the same time, this ratio increases with change in $\omega_L$ (see inset (b2) of Fig. 6 (b)). We note that such increase in intensity ratio for change in incident photon energy is suggested by Allen et al., [32] for $3d$ oxides invoking self-trapped exciton (orbitons) which results in strong multiphonon Raman scattering. But within their model, second-order to first-order intensity ratio is expected to be temperature independent. Interestingly, second to first-order ratio (i.e. $I_{16+17}/I_{10+11}$) for the band centred near 1150 cm$^{-1}$ is quite high ($\sim$ 0.35) and remains almost constant in the entire temperature range (see Fig. 5a). This anomalous temperature and incident photon energy dependence of these modes clearly suggest that these high energy modes are not pure two phonons but have contributions from orbital degrees of freedom.

Figure 6 (a, b) shows the Raman spectra as a function of different incident photon energy (i.e. $\omega_L$ = 633 (1.96 eV), 532 (2.33 eV) and 325 (3.81 eV) nm). Following observations can be made: (i) mode S2 and S4 show a weak signal with UV excitation and strong resonance in the visible range (see inset (a1) of Fig. 6 (a)). We note that Jahn-Teller gap ($\Delta_{CF}$) in iridate based systems is of the order of $\sim$ 2.5 eV, therefore resonance of modes S2 and S4 in visible range suggests their potential coupling with Jahn-Teller. (ii) Higher energy modes (i.e. S10 and S11) associated with oxygen vibrations show resonance in the UV region whereas mode S12 shows resonance in the visible range (see inset (a2)). (iii) Two-phonon bands with centering near 1150 cm$^{-1}$ and 1450 cm$^{-1}$ show resonance behaviour in the visible range (see inset (b1)). The two phonon Raman band near 1450 cm$^{-1}$ (i.e. S18 and S19) show similar matrix element as that of their first-order (i.e. S12 and S13) as both resonate in the visible range (see inset (a2) and (b1)). What is most interesting and surprising is the opposite resonance behaviour of first-order modes near      550 cm$^{-1}$ (i.e. S10 and S11) and their corresponding second-order modes near 1150 cm$^{-1}$ (i.e. S16 and S17), (see inset (a2) and (b1)) again suggesting that these higher energy modes can not be pure two-phonon modes.

Now, the natural question is, why the intensity ratio (i.e. $I_{16+17}/I_{10+11}$) is constant with temperature and these modes show opposite resonance behaviour w.r.t. their first-order counterparts. We have tried to understand the anomalies in these high frequency Raman bands via invoking orbital excitations, which are collective or local excitations of the orbitals. Theoretically, there are two different approaches proposed in case of 3$d$ oxides, first by Brink [43], where the coupling between orbiton-phonon mediated by strong electron-phonon coupling leads to satellite structure in the phonon spectrum that has the mixed character of phonon and orbitons. In the second picture, proposed by Allen et al., [32] rearrangement of oxygen in the octahedra leads to self-trapped orbital exciton, where Franck-Condon process leads to intense Raman active multiphonon bands with almost same intensity as their first-order counterpart and it increases as one move towards resonance. Looking at the temperature and $\omega_L$ dependent behaviour of these high energy bands, they clearly reflect that they are not alone second-order phonon modes, and definitely have significant contribution from orbital degrees of freedom. The symmetry analysis also points that phonons and orbital excitations couple strongly in line with our observations. Two quasi-particle excitations couples strongly if they have similar energy and associated symmetry with these excitations is same. The high energy phonon modes associated with IrO$_6$ octahedra and showing anomalous behaviour are in the energy range of ~ 200 meV and the associated symmetries are $A_g + B_g$. The possible symmetry of transition between different levels of $t_{2g}$ manifolds are given by the direct product of the irreducible representations: $E_g \otimes E_g = A_{1g} + A_{2g} + B_{1g} + B_{2g}$, $B_{2g} \otimes E_g = E_g$ and $B_{2g} \otimes B_{2g} = A_{1g}$. The allowed symmetry of transition between $t_{2g}$ levels are given by $t_{2g} \otimes t_{2g} = A_{1g} + E_g + T_{1g} + T_{2g}$. Therefore, group theoretical symmetry allow these quasi-particle excitations to couple, and the splitting of $t_{2g}$ manifold driven by phononic degrees of freedom is expected to be of this order ($\delta_{CF} \sim 200$ meV). We note that the strength of the SOC in this system is $\sim 350$ - $400$ meV, and non-cubic crystalline field is of comparable strength and this clearly suggests the strong competition between these two controlling parameter for deciding the complex ground state of this system. As mentioned earlier, if the systems is fully governed by SOC then $J_{eff} = 0$ state will be the ground state, however the strong competition from the non-cubic crystalline field suggest that $J_{eff} = 0$ picture needs to be revisited. Our results clearly points toward the intricate role of phononic and orbital degrees of freedom in these systems and they should be treated at par with other crucial parameter for unravelling the underlying properties of these systems.

## Acknowledgments

PK thanks the Department of Science and Technology (DST), India, for the grant under INSPIRE Faculty scheme and Advanced Material Research Center, IIT Mandi, for the experimental facilities. The authors at Dresden thanks DFG program for financial support.

**Table-I:** List of the experimentally observed phonon frequencies at 80K, fitting parameters, fitted using equations as described in the text and calculated phonon frequencies for both experimental as well as theoretical lattice parameters. Units are in cm$^{-1}$. Mode assignment (with displacement of atoms involved) has been done based on our DFT based calculations. Modes S4-S15 are associated with the vibrations of oxygen octahedral (IrO$_6$) and modes S1-S3 are associated with prominent displacement of Ba atom.

| Mode Assignment | Exp. $\omega$ | Fitted Parameters | | | | DFT - $\omega$ | |
| --- | --- | --- | --- | --- | --- | --- | --- |
| | | $\omega_0$ | A | $\Gamma_0$ | C | Exp. Lat. const. | Theo. Lat. const. |
| S1-B$_g$ (Ba, Y/ Ir) | 102.5 | 103.3 ± 0.2 | -0.21 ± 0.02 | 3 ± 0.7 | 1.26 ± 0.1 | 95.6 | 103.7 |
| S2-B$_g$ (Ba, O) | 106.5 | 106.2 ± 0.1 | 0.16 ± 0.01 | 0.9 ± 0.1 | 0.15 ± 0.01 | 99.1 | 107.7 |
| S3-B$_g$ (Ba, O) | 130 | 131.1 ± 0.3 | -0.08 ± 0.06 | 11.3 ± 0.4 | 0.51 ± 0.08 | 127.2 | 128.7 |
| S4-A$_g$ (IrO$_6$) | 216.8 | 218.4 ± 0.1 | -0.91 ± 0.04 | 5.5 ± 0.2 | 1.19 ± 0.06 | 221.8 | 226.2 |
| S5-B$_g$ (IrO$_6$) | 227.6 | 230.2 ± 0.2 | -1.35 ± 0.08 | 11.3 ± 0.3 | 1.78 ± 0.1 | 223.9 | 228.5 |
| S6-B$_g$ (IrO$_6$) | 289.5 | 286.3 ± 0.6 | 4.02 ± 0.3 | 31.5 ± 1.3 | 2.49 ± 0.5 | 285 | 294.3 |
| S7-A$_g$ (IrO$_6$) | 324.2 | 322.5 ± 0.6 | 2.38 ± 0.3 | 31.1 ± 1.4 | 2.59 ± 0.7 | 326.6 | 333.7 |
| S8-A$_g$ (IrO$_6$) | 349.7 | 347.8 ± 0.1 | 1.75 ± 0.06 | 0.23 ± 0.2 | 2.17 ± 0.1 | 337.8 | 346.3 |
| S9-B$_g$ (IrO$_6$) | 397.6 | 398.1 ± 0.1 | -0.46 ± 0.05 | 12.1 ± 0.3 | 2.58 ± 0.2 | 371 | 381.8 |
| S10-A$_g$ (IrO$_6$) | 561.7 | 572.5 ± 0.3 | -11.13 ± 0.2 | | | 554.6 | 617.5 |
| S11-A$_g$ (IrO$_6$) | 580.4 | 589.7 ± 0.5 | -9.56 ± 0.5 | | | 561.7 | 621.5 |
| S12-B$_g$ (IrO$_6$) | 711 | 726.9 ± 1.5 | -17.26 ± 1.4 | | | 568.5 | 659.8 |
| S13-B$_g$ (IrO$_6$) | 734.7 | 760.1 ± 1.9 | -26.2 ± 1.6 | | | 588.8 | 667 |
| S14-B$_g$ (IrO$_6$) | 773.5 | 798.6 ± 1.1 | -25.44 ± 1 | | | 685.2 | 756.4 |
| S15-A$_g$ (IrO$_6$) | 785.5 | 805.2 ± 0.7 | -19.95 ± 0.7 | | | 732.1 | 788.6 |
| S16 | 1113 | | | | | | |
| S17 | 1165 | | | | | | |
| S18 | 1420 | | | | | | |
| S19 | 1483 | | | | | | |

## REFERENCES:

[1] F. Wang and T. Senthil, Phys. Rev. Lett. **106**, 136402 (2011).

[2] X. Wan, A. M. Turner et al., Phys. Rev. B **83**, 205101 (2011).

[3] I. I. Mazin, H. O. Jeschke et al., Phys. Rev. Lett. **109**, 197201 (2012).

[4] G. Cao, T. F. Qi et al., Phys. Rev. Lett. **112**, 056402 (2014).

[5] G. Khaliullin, Phys. Rev. Lett. **111**, 197201 (2013).

[6] O. N. Meetei, W. S. Cole et al., Phys. Rev. B **91**, 054412 (2015).

[7] B. J. Kim, H. Jin et al., Phys. Rev. Lett. **101**, 076402 (2008).

[8] B. J. Kim, H.Ohsumi et al., Science **323**, 1329 (2009).

[9] J. G. Rau, E. K. H. Lee and H. Y. Kee, Annu. Rev. Condens Matter Phys. **7**, 195 (2016).

[10] M. Bremholm, S. E. Dutton et al., J. Solid State Chem. **184**, 601 (2011).

[11] T. Dey, A. Maljuk et al., Phys. Rev. B **93**, 014434 (2016).

[12] J. Terizc, H. Zhang, et al., *Phys. Rev. B* **96,** 064436 (2017).

[13] B. Ranjbar, E. Reynolds et al., Inor. Chem. **54**, 10468 (2015).

[14] S. Bhowal, S. Baidya et al., Phys. Rev. B **92**, 121113 (2015).

[15] L. H. Hall, Group theory and symmetry in chemistry (McGraw Hill, New York, 1969).

[16] M. Wakeshima, D. Harada and Y. Hinatsu, J. Alloys and Compound **287**, 130 (1999).

[17] N. A. Bogdanov, V. M. Katukuri et al., Nature Commun. **6**, 7306 (2015).

[18] D. Vanderbilt, Phys. Rev. B **41,** 7892 (1990).

[19] P. Giannozzi  et al.,  J. Phys: Cond. Matter **21,** 395502 (2009).

[20] J. P. Perdew, et al., Phys. Rev. Lett. **100,** 136406 (2008).

[21] P. Giannozzi, et al., Phys. Rev. B **43,** 7231 (1991).

[22] H. J. Monkhorst, et al., Phys. Rev. B **13,** 5188 (1976).

[23] I. Thumm, U. Treiber, and S. K. Sack, J. Solid State Chem. **35**, 156 (1980).

[24] D. L. Rousseau, R. P. Bauman and S. P. S. Porto, J. Raman Spectro. **10**, 253 (1981).

[25] J. Moreira, A. Almeida et al., Phys. Rev. B **76**, 174102 (2007).

[26] P. Kumar et al., AIP Advances **5**, 037135 (2015).

[27] M. Balkanski, R. F. Wallis and E. Haro, Phys. Rev. B **28**, 1928 (1983).

[28] P. Kumar et al., Phys. Rev. B **85**, 134449 (2012).

[29] E. Granado, A. Garcia et al., Phys. Rev. B **60**, 11879 (1999).

[30] M. Cardona and T. Ruf, Solid State Commun. **117**, 201 (2001).

[31] E. Saitoh et al., Nature **410**, 180 (2001).

[32] P. B. Allen and V. Perebeinos, Phys. Rev. Lett. **83**, 4828 (1999).

[33] R. Kruger, B. Schulz et al., Phys. Rev. Lett. **92**, 093203 (2004).

[34] L. M. Carron and A. de Andres, Phys. Rev. Lett. **92**, 175501 (2004).

[35] C. Ulrich, A. Gossling et al., Phys. Rev. Lett. **97**, 157401 (2006).

[36] P. Kumar et al., J. Phys. Cond. Matter **22**, 115403 (2010).

[37] K. I. Kugel and D. I. Khomski,   Sov. Phys. Usp. **25**, 231 (1982).

[38] E. M. Plotnikova, et al., Phys. Rev. Lett. **116**, 106401(2016).

[39] C. Svoboda, et al., Phys. Rev. B **95**, 014409 (2017).

[40] H. Gretarsson, et al., Phys. Rev. Lett. **116**, 136401 (2016).

[41] B. Yuan, et al., Phys. Rev. B **95**, 235114 (2017).

[42] M. Cardona, in Light scattering in solids II, edited by M. Cardona and G. Guntherdot, Topics in applied physics, Vol.50 (springer-verlag, Berlin, 1982).

[43] J van den Brink, Phys. Rev. Lett. **87**, 217202 (2001).

**FIGURE CAPTION:**

**FIGURE 1:** (Color online) Schematic showing the $Ir^{5+}$ $d$-orbitals under different perturbations **(i)** degenerate $d$-orbitals for the free Ir ion **(ii)** $d$-orbitals splits into $t_{2g}$ and $e_g$ under cubic crystalline field ($\Delta_{CF}$) **(iii)** compression of the octahedra along z-axis and/or strong anharmonic phonons causes non-cubic crystalline field ($\delta_{CF}$) and led to orbital singlet ($d_{xy}$) to be lowest in energy **(iv)** SOC dominated splitting of $t_{2g}$ level which lead to an orbital doublet ($d_{xz}, d_{yz}$) to be lowest in energy.

**FIGURE 2:** (Color online) Raman spectra of $Ba_2YIrO_6$ at 80K. Solid thin lines are the fit of individual modes, and solid thick line shows the total fit to the experimental data. Inset (b) shows the spectrum at 300K and inset (a) shows the temperature evolution of mode S10 and S11 showing the spectral weight transfer from mode S10 to S11 as temperature is lowered.

**FIGURE 3:** (Color online) Temperature evolution of the self-energy parameters of modes S2, S4, S8 and S9. Solid lines are the fitted curve as described in the text. Pictures at the center show the graphic representation of the eigenvectors for the modes S2, S4, S8 and S9.

**FIGURE 4:** (Color online) Temperature evolution of the self-energy parameters of modes S10-S12 and S15. Solid lines are the fitted curve as described in the text. Graphic representation of the eigenvector of modes S10-S12 and S15.

**FIGURE 5:** (Color online) **(b, c, e and f)** Temperature evolution of the peak frequency (filled circle) and FWHM (empty circle) of modes S16-S19. **(a) and (d)** Intensity ratio of the Raman band centered around 1150 cm$^{-1}$ (i.e. S16 and S17) and 1450 cm$^{-1}$ (i.e. S18, S19) w.r.t. their first-order counterpart i.e. S10, S11 and S12, S13, respectively.

**FIGURE 6: (a)** Room temperature Raman spectra of first-order phonon modes as a function of different incident photon energies (i.e. $\omega_L$ = 633 (1.96 eV), 532 (2.33 eV) and 325 (3.81 eV) nm). Inset (**a1 and a2**) shows the resonance profile of mode S2, S4 and S10, S12, respectively. **(b)** High energy Raman bands as a function of different incident photon energies. Inset (**b1**) shows the resonance profile of normalized intensity of bands near 1150 cm$^{-1}$ (i.e. $I_{16+17}$) and 1450 cm$^{-1}$ (i.e. $I_{18+19}$). Inset (**b2**) shows the resonance profile of the normalized intensity ratio of high energy Raman bands w.r.t. their first-order counterparts.

**FIGURE 1:**

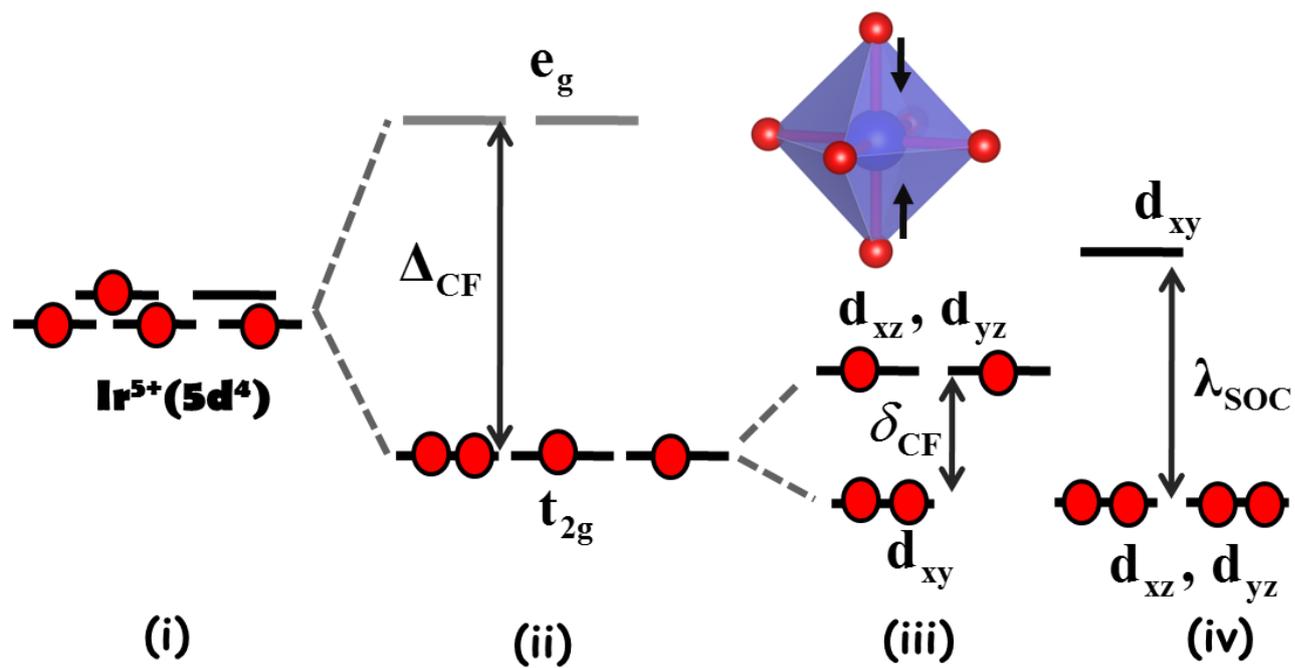

$\Delta_{CF}$

$e_g$

$\mathbf{Ir^{5+}(5d^4)}$

$t_{2g}$

$\mathbf{d_{xz}, d_{yz}}$

$\delta_{CF}$

$\mathbf{d_{xy}}$

$\mathbf{d_{xy}}$

$\lambda_{SOC}$

$\mathbf{d_{xz}, d_{yz}}$

(i)          (ii)          (iii)          (iv)

**FIGURE 2:**

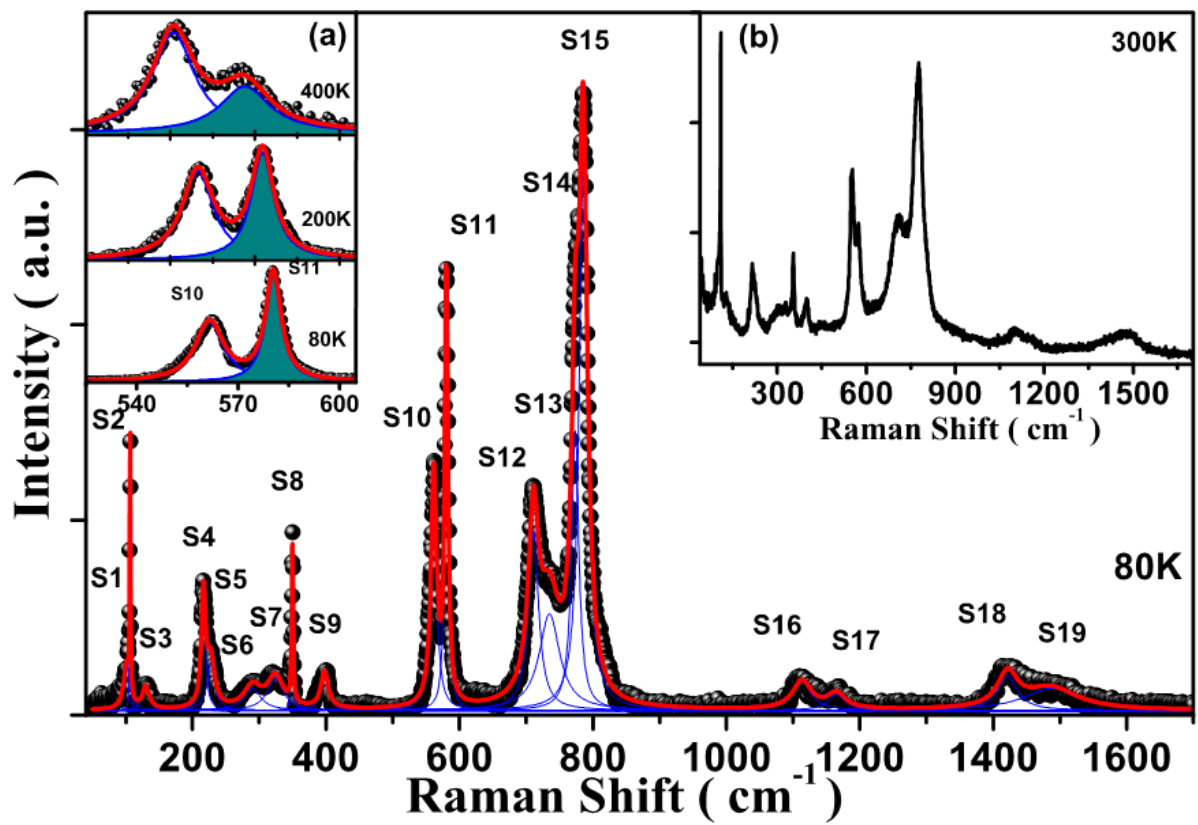

**FIGURE 3:**

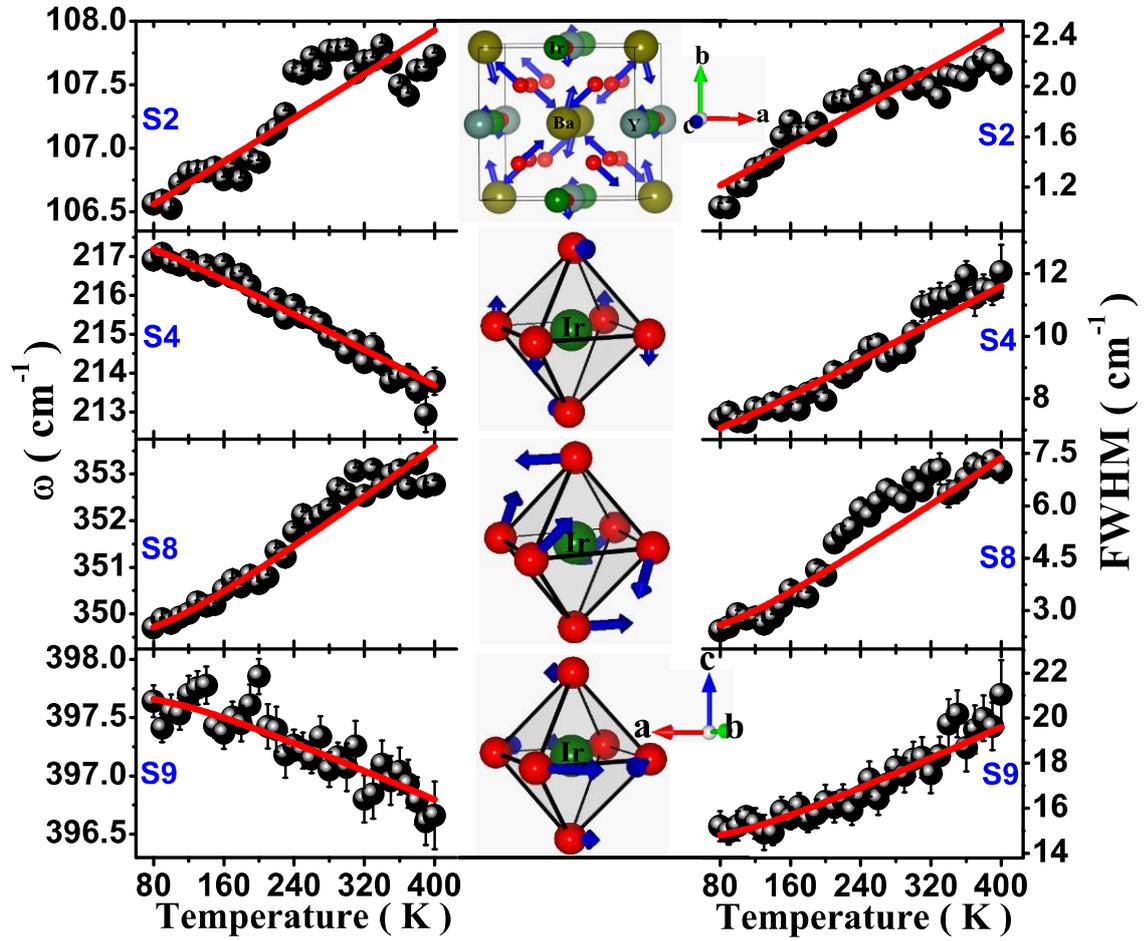

**FIGURE 4:**

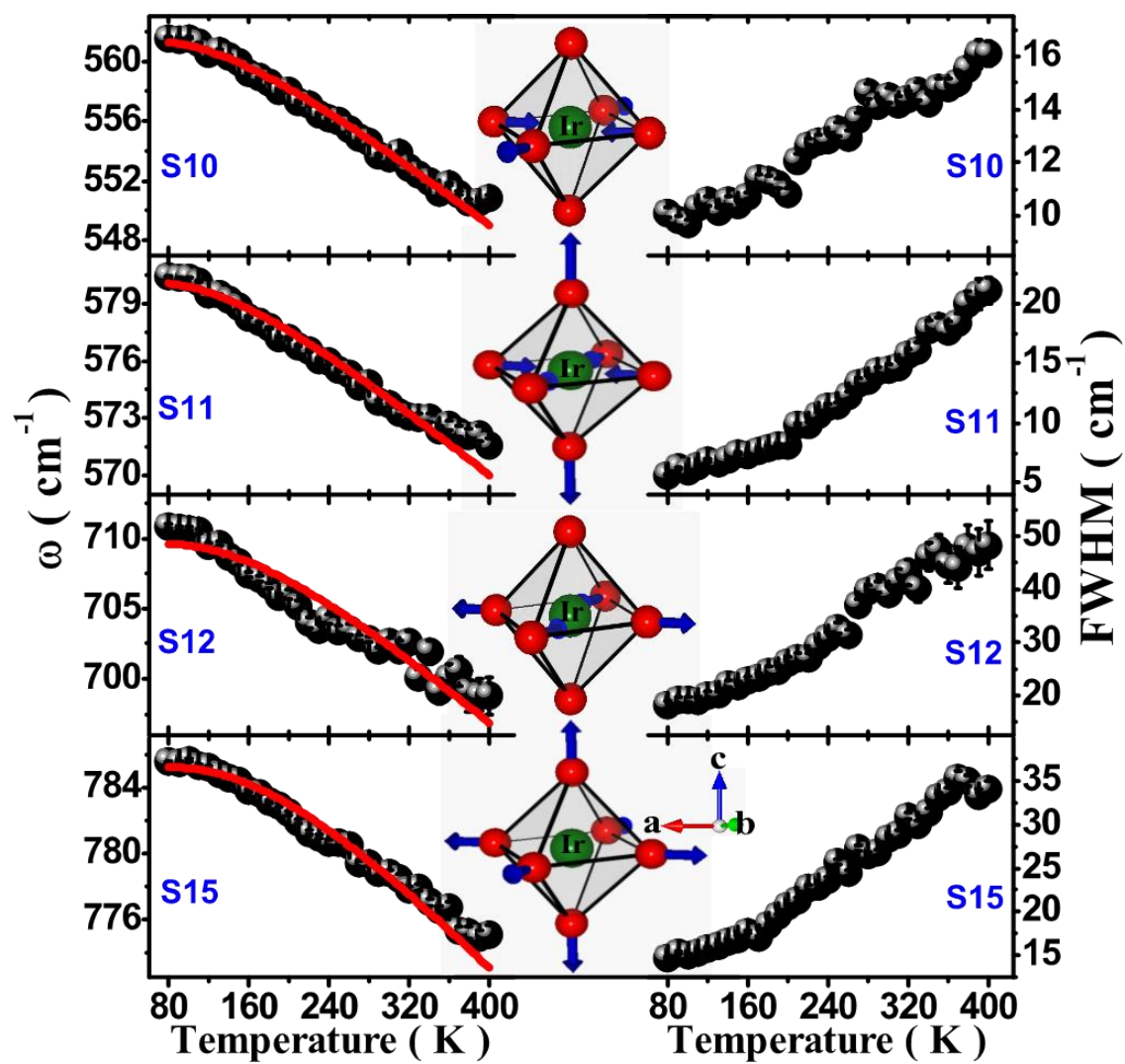

**FIGURE 5:**

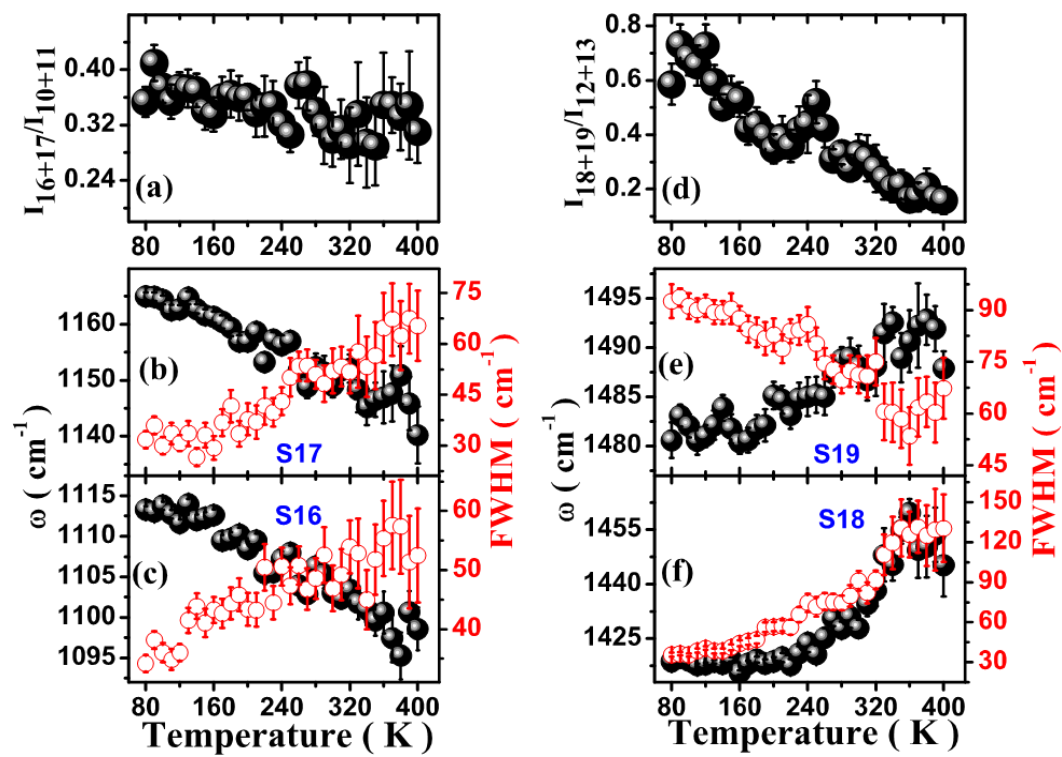

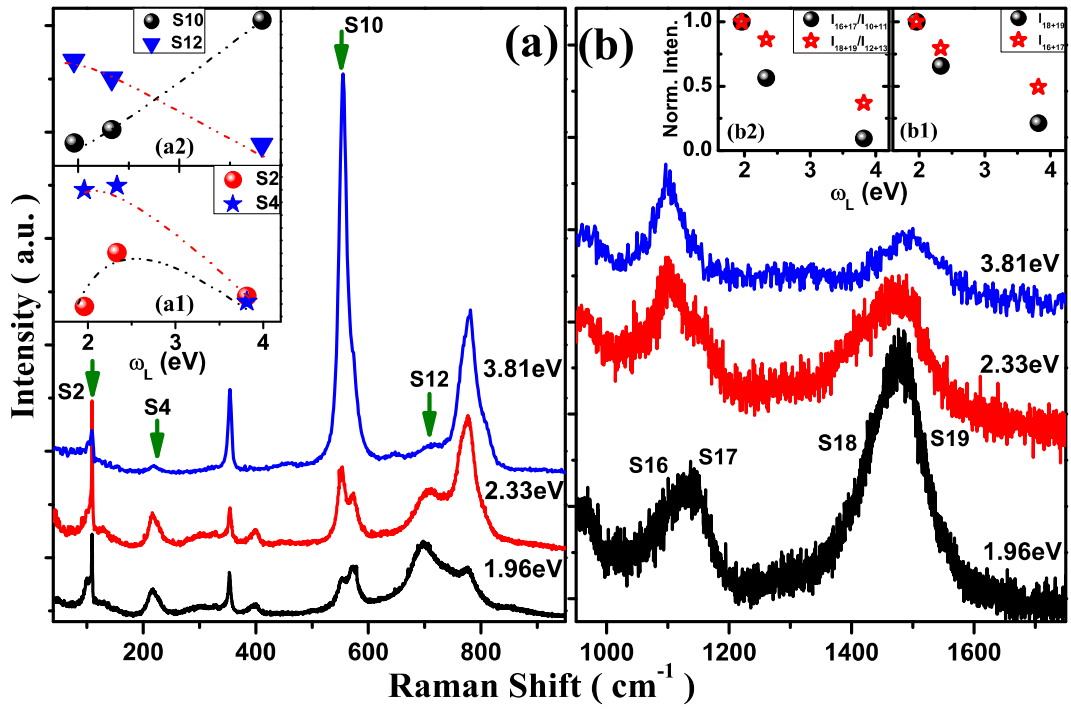